\documentstyle[12pt,physrev]{article}

\begin{document}

\begin{center}

\thispagestyle{empty}

{\large {\bf Medium corrections to the tensor interaction in  nuclei:
Energy difference of $T$=1 and $T$=0 $J^{\pi}$=$0^{-}$
states in $^{16}\mbox{O}$}}

\vspace{0.2in}

D. C. Zheng$^{(1,2)}$, L. Zamick$^{(3)}$, and H. M\"uther$^{(4)}$\\

\end{center}

\vspace{0.2in}
\begin{small}

\hspace*{-0.2in}
{\it $^{(1)}$Institute for Nuclear Theory, University of Washington,
                Seattle, Washington 98195}

\hspace*{-0.4in}
{\it $^{(2)}$Department of Physics, University of Arizona, Tucson, Arizona
                85721}

\hspace*{-0.4in}
{\it $^{(3)}$Department of Physics and Astronomy, Rutgers University,
     Piscataway, New Jersey 08855}

\hspace*{-0.4in}
{\it $^{(4)}$Institut f\"{u}r Theoretische Physik der
        Universit\"{a}t T\"{u}bingen, D--7400 T\"{u}bingen, Germany}

\end{small}

\vspace{0.5in}

\begin{abstract}
\noindent
It has been shown in the past that in a shell model calculation,
restricted to
$1\hbar\omega$ $1p$-$1h$ configurations, the energy splitting
between the lowest $J^{\pi}$=$0^{-}$, $T$=0 and $J^{\pi}$=$0^{-}$, $T$=1
states in $^{16}\mbox{O}$ is mainly due to the tensor component of
the nucleon-nucleon interaction.
In this work, we extend the calculation to include
$3\hbar\omega$ $2p$-$2h$ and $3p$-$3h$ configurations. We use
realistic G matrices derived from various One-Boson-Exchange potentials
taking into account the effects of the Dirac-Brueckner-Hartree-Fock
approach, i.e., a modification of the Dirac spinors for the nucleons in
the medium. It is found that the splitting is rather sensitive to the
long-range components of the isovector tensor force. The splitting is
reduced considerably by including $2p$-$2h$ configurations, in
particular.
\end{abstract}

\pagebreak

\section{Introduction}
\noindent
Besides the two-nucleon system, also the structure of nuclei is a very
important source
for our understanding of the nucleon-nucleon (NN) interaction.
Previously we have studied selected $0p$ shell nuclei, focusing on
the properties which are sensitive to various components
(e.g. spin-orbit, tensor, {\it etc.}) of the NN
interaction \cite{ann,zzm,spe,zz}. We found that, for shell model
calculations in a small model space, an effective interaction is needed
with a strong spin-orbit component and/or a weak tensor component in
order to reproduce the experimental data. Such an effective interaction
can be derived from a realistic NN interaction by including core polarization
terms (which simulate the effects of a larger model space) and/or taking
into account the effects of the Dirac-Brueckner-Hartree-Fock (DBHF) approach
by evaluating the matrix elements of the One-Boson-Exchange (OBE)
potentials for nucleon Dirac spinors in the medium, which are
characterized by an effective mass for the nucleon smaller than the
bare mass. This implies that the NN interaction for two nucleons in the
medium is different from the NN in the vacuum since the Dirac spinors
used to evaluate the matrix elements of the NN interaction in the
medium are modified as compared to the Dirac spinors in the vacuum. It
is one of our aims to search for observables in nuclear structure
calculations, which give a clear indication whether such medium
modifications of a realistic NN interaction are required to obtain a
microscopic understanding of nuclear structure data.

In this work, we continue our effort towards a better understanding
of the NN interaction in nuclei through studying nuclear properties by
focusing on the energy splitting between $J^{\pi}$=$0^{-}$ $T$=0
and $J^{\pi}$=$0^{-}$ $T$=1 states in $^{16}\mbox{O}$.
We recall early works by Blomqvist and Molinari \cite{blom}, Millener
and Kurath \cite{mk}, and Barrett \cite{barrett},
in which it is pointed out that this energy splitting
is due mainly (but not exclusively) to the tensor interaction.
Thus a careful study of the splitting
could help us to determine the strength of the tensor interaction.

We will again use relativistic versions of the Bonn OBE potentials \cite{zzm}
allowing for the DBHF effects discussed above. As a first step we will perform
the calculation in a small space, in which the $0^{-}$ states are
treated as $1\hbar\omega$ one-particle, one-hole ($1p$-$1h$) states.
We will then extend our model space to include
$3\hbar\omega$ $3p$-$3h$ configurations with three particles
in the $1s$-$0d$ shell and three holes in the $0p$ shell.
Finally, we will perform a calculation allowing
also $2p$-$2h$ configurations with two particles in the
$1s$-$0d$ and $1p$-$0f$ shells and two holes in the $0s$ and $0p$ shells.
This comparison of results obtained in various model spaces should help
us to study the various ingredients leading to our most complete
estimate for the investigated energy splitting.

After this short introduction, section 2 will present some details on
the various hamiltonians which we consider. The detailed discussion of
our results is given in section 3, and section 4 contains our main
conclusions.

\section{Details of the Calculations}

\noindent
In order to determine the matrix elements for the effective interaction
of two nucleons in a nuclear medium we consider the OBE potentials A
and C defined in table A.2 of \cite{mach}. Since the parameters for
these potentials have been determined employing the Thompson equation,
one of the 3-dimensional approximations to the Bethe-Salpeter equation,
in the fit of the NN scattering data, we also evaluate as a first step
the scattering matrix $T$ in this approximation. This has been done
assuming for the Dirac spinors of the nucleons the solution of the free
Dirac equation (hereafter identified by an effective mass $m^\ast = m$)
and by considering Dirac spinors for which the ratio of small to large
component is characterized by an effective mass $m^\ast$ = 729 MeV,
which is a typical result obtained in DBHF calculations \cite{mmb}. The
resulting matrix elements for $T$ are identified as matrix elements
between non-relativistic plane wave states and matrix elements in an
oscillator basis are derived from $T$ in the conventional way. For the
basis of oscillator states we consider an oscillator length $b$=1.72
fm, a value which defines a basis appropriate for $^{16}\mbox{O}$.
Within this
oscillator basis, the conventional techniques can be used to derive the
$G$ matrix form $T$ by solving
\begin{equation}
G = T + T \left[ \frac{Q}{\omega - QtQ} - \frac{1}{\omega - t}\right] G
\end{equation}
Here $\omega$ represents the starting energy which we have chosen as
$\omega$ = -30 MeV. The kinetic energy of the intermediate particle states
is given by $t$ and $Q$ is a Pauli operator
appropriate for the model space to be used in the shell model. We have
used an operator $Q$ which suppresses intermediate states with one
nucleon in $0s$ or $0p$ states and those with two nucleons in
$1s$-$0d$ and $1p$-$0f$ shells as such configurations will be taken
into account explicitly in our most complete shell model calculation.

In addition, one has to define the kinetic energy for nucleons
characterized by Dirac spinors discussed above. For a nucleon with
momentum $\mbox{\boldmath p}$ this yields
\begin{equation}
t_{\rm rel} = \frac{\mbox{\boldmath p}^2 + mm^\ast}{\sqrt{\mbox{\boldmath
p}^2 + {m^\ast}^2}} - m\; .
\end{equation}
In order to evaluate the matrix elements for this operator in a basis
of harmonic oscillator states, it is useful to expand this expression
\begin{equation}
t_{\rm rel} = \frac{2mm^*-m^2}{{m^*}^2} \left(\frac{\mbox{\boldmath $p$}^{2}}
{2m} \right) + \frac{3m^3-4m^2 m^*}{2{m^*}^4}
\left( \frac{\mbox{\boldmath $p$}_{i}^{2}}{2m} \right)^2
+ \cdots\; . \label{eq2}
\end{equation}
An expansion up to terms of second order in $\mbox{\boldmath p}^2$ turned
out to be sufficient \cite{mmb}.

In order to see how the shell model calculations are affected by
different components of the NN interaction,
we will also use a schematic interaction
which consists of three main components \cite{ann}:
\begin{equation}
V_{\rm sche} = V_{\rm c}+xV_{\rm so}+yV_{\rm t},  \label{eq3}
\end{equation}
where c=central, so=two-body spin-orbit, and t=tensor. For $x$=$y$=1,
the matrix elements for $V_{\rm sche}$ are in approximate agreement with
those of the non-relativistic OBE potential A of \cite{mach}.
In this schematic calculation we
use the non-relativistic form of the kinetic energy, i.e. the first
term in the expansion of eq.(\ref{eq2}) considering $m^\ast = m$.
The interaction is defined in terms of the relative matrix elements:
\begin{equation}
\langle nl|{\cal V}_{c}|n'l'\rangle_{ST{\cal J}}
= \left[ -8.6(1+0.1\vec{\sigma}\cdot\vec{\sigma}')\delta_{l,0}
  +1.7\delta_{l,1}\delta_{S,0}-0.6\delta_{l,2}\delta_{S,1}\right]
      \delta_{l,l'},
\end{equation}
\begin{equation}
\langle nl|{\cal V}_{so}|n'l'\rangle_{ST{\cal J}}
 = -0.375 \delta_{l,1}\delta_{l',1}\delta_{S,1}[{\cal J}({\cal J}+1)-4],
                    \label{so}
\end{equation}
\begin{equation}
\langle nl|{\cal V}_{t}|n'l'\rangle_{ST{\cal J}}
= [-5.5\theta(n'-n)-3\theta(n-n'-1)]\delta_{l,0}\delta_{l',2}
    + 1.8\delta_{l,l'}(\delta_{l,1}-\delta_{l,2})\tilde{S}_{12}.
              \label{tens}
\end{equation}
We refer the reader to Ref.\cite{ann} for further discussion of
this interaction.

It should be emphasized that
we do not insert empirical single-particle
energies into the calculation.
Rather they are implicitly generated by the kinetic energy and the same
interaction that is
used to calculate the matrix elements of the residual interaction.

\section{Results and Discussion}

\noindent
In Tables I and II we present the results for the energies of the lowest
$J^{\pi}$=$0^{-}$, $T$=0 and $J^{\pi}$=$0^{-}$, $T$=1 states in $^{16}\mbox{O}$
and focus our interest on the energy difference of these two states.
This means that we are mainly studying the isovector component of the
NN interaction. The results displayed in Table I were obtained for the
schematic interaction $V_{\rm sche}$ of Eq.(\ref{eq3}) and those in Table II
were obtained employing the OBE potentials A and C discussed above.
Experimentally the excitation energies of these two states
are 10.952 MeV and 12.797 MeV, respectively,
so the energy splitting between them is 1.845 MeV.

We first performed a $1p$-$1h$ calculation.
In this simple calculation, the dominant configuration is
$(1s_{1/2}\; 0p_{1/2}^{-1})^{0^{-}}$, but there is also a
small admixture of $(1d_{3/2}\; 0p_{3/2}^{-1})^{0^{-}}$
in the wave function.
With a purely central interaction ($x$=$y$=0),
as shown in Table I, the splitting $\Delta E$ is very small (0.017 MeV).
Also the spin-orbit term of the schematic interaction seems to have
a very small isovector component in our model space and therefore the
splitting remains small (0.046 MeV) even after the spin-orbit
contribution has been added to the central one ($x$=1, $y$=0).
The inclusion of the spin-orbit term, however, lowers the
excitation energy of both the $T$=0 and $T$=1 states appreciably. This
is mainly due to the fact that the inclusion of a spin-orbit term
removes the degeneracy between the $p_{3/2}$ and $p_{1/2}$ states (as
well as the one between the $d$ states) \cite{ann} and therefore one
obtains a smaller excitation energy for the $(1s_{1/2}\;
0p_{1/2}^{-1})^{0^{-}}$ configuration.

When the tensor interaction is introduced ($x$=0, $y$=1), a
substantial splitting of 3.144 MeV is obtained, confirming
the early ideas of Blomqvist and Molinari \cite{blom},
Millener and Kurath \cite{mk}, and Barrett \cite{barrett}. Of course
this also demonstrates that the tensor force has a strong isovector
component, which is quite obvious from the main source of this tensor
force, the One-Pion-Exchange, and can also be seen from the
parameterization in \cite{ann}.
When the full interaction ($x$=$y$=1) is used, we get
a splitting of 2.932 MeV, which is close to that for the
central plus tensor interaction.
We further note that when we vary the tensor strength $y$,
the isospin splitting $\Delta E$ is, to an excellent approximation,
linear in $y$.

We now improve the calculation by allowing more configurations.
We allow not only $1\hbar\omega$ $1p$-$1h$ but also
$3\hbar\omega$ $3p$-$3h$ configurations in our model space for the
$J^{\pi}$=$0^{-}$ states, while at the same time, allowing
$2\hbar\omega$ $2p$-$2h$ configurations in the ground state.
When these additional admixtures are introduced,
we get an overall increase for the calculated excitation energies
but the splitting $\Delta E=E(J^{\pi}=0^-_1,T=1)-E(J^{\pi}=0^-_1,T=0)$
remains almost unchanged.
{\it A priori} we might have expected changes, because $3p$-$3h$,
highly-deformed intruder states tend to come down very low
in energy. Our calculation does not show this happening.
However, a proper description
of the intruders would probably require a much larger shell model space
than the one we are using.

In order to discuss our results a bit more in detail and to show the
connection to other approaches we will discuss the various
contributions in terms of the diagrams displayed in Fig.1. The diagrams
displayed in Figs.1a)-1d) represent the effects of the residual
interaction on the 1p-1h excitation. Diagrams 1a) and 1b) exhibit only
intermediate 1p-1h excitations. These terms are taken into account in a
standard Tamm-Dancoff calculation and are also present in our 1p-1h
shell model calculation. The diagram displayed in Fig.1c) exhibits the
contribution of second order in the residual interaction, which would
be taken into account employing the RPA approximation. In the shell
model scheme, the diagram 1c) is taken into account by allowing for
$3p$-$3h$ excitations in the model space. While, however, RPA calculations
employing fixed single-particle energies, account for diagram 1c) only,
a shell model calculation will include additional diagrams, among
others also those of figure 1e). These diagrams are to be interpreted
as modifications of the single-particle energies for particle- and
hole-states. Such modifications should have the same effect on the $T=0$
states as on those with $T=1$.  Indeed we find that the inclusion of
$3p$-$3h$ configurations enhances the excitation energies without
changing the energy splitting. This is different from results obtained
in RPA where the modifications of single-particle energies exhibited in
Fig.1e) are ignored.

We then extended the calculation to include the $3\hbar\omega$
$2p$-$2h$ states. Here, surprisingly, we get {\it large} effects.
With the full
schematic interaction ($x$=$y$=1), we find in Table I
that the splitting of $T$=1 and
$T$=0 states decreases from the $1p$-$1h$ result of
2.932 MeV to 2.002 MeV, the latter being very close to experiment (1.845 MeV).

Translated into the language of diagrams, the inclusion of $2p$-$2h$
configurations in the shell model is represented by Fig.1d) and 1f).
Both contributions are ignored in standard RPA. The fact that the
inclusion of $2p$-$2h$ configurations modifies the energy splitting in
a significant way shows that terms like Fig.1d) play an important
role. Within RPA such terms could be taken into account by
renormalizing the residual interaction, accounting for the effects of
the so-called induced interaction \cite{wim}. The splitting $\Delta E$
becomes at least
30\% smaller due to this graph --- indeed we can explain
the data. We can thus claim that the graph of Fig.1d)
represents a key medium correction. It leads to a weaker effective tensor
interaction, responsible for the energy splitting.

Note that Fig.1d)
corresponds to the so-called bubble-diagram, well known in the
renormalization of effective shell model interaction. As in the case of
the bubble-diagram in the shell model interaction \cite{som}, we also
find here that effects of 1d) are dominated by the tensor component of
the NN interaction. If we only have the central interaction, the value
of $\Delta E$ in Table I changes from 0.017 MeV to 0.109 MeV as
we go from $1p$-$1h$ to the ``full''
$1p$-$1h$+$2p$-$2h$+$3p$-$3h$. We get a slight enhancement.
However, when we have both the central and the tensor
interactions, $\Delta E$ changes from
3.144 MeV to 2.238 MeV.

When we include the two-body spin-orbit interaction, the absolute
energies of the $J^{\pi}$=$0^-$ states go down. This is not
surprising because the $0p_{1/2}$
orbit gets pushed up closer to the $1s_{1/2}$ orbit, thus causing
the excitation energy of the state $(1s_{1/2}\; 0p_{1/2})^{J^{\pi}=0^-}$
to be lower. Note that although the one-body spin-orbit
interaction does not contribute to the splitting $\Delta E$,
there is a small contribution from the two-body spin-orbit
interaction. But again, it is only when the tensor interaction
is present in evaluating Fig.1d does the value of $\Delta E$
undergo a considerable
reduction.

The trends for Bonn A and Bonn C are the same as that for the
schematic interaction in the sense that the
$1p$-$1h$ + $3p$-$3h$  calculation gives almost the same
isospin splitting as does the $1p$-$1h$ calculation, but
the inclusion of the $3\hbar\omega$ $2p$-$2h$ configurations
decreases the splitting considerably.

There are other points of interest in Table II,
especially the comparison of the results of Bonn A and Bonn C.
Now Bonn C produces a larger D-state admixture in the deuteron
than Bonn A does (5.6\% vs 4.4\%) \cite{mmb}.
Thus, the tensor interaction is stronger in potential C than in OBEP A. If
the splitting would be explained only by the tensor interaction, one
would expect a larger value of $\Delta E$ for the potential C than for
A. The results of Table II, however, show just the opposite:
the value of $\Delta E$ is smaller for Bonn C than for
Bonn A. In the case of $m^{*}$=939 MeV, the values in the
$1p$-$1h$ calculation are 2.72 MeV for OBEP C and 3.08 MeV for the
potential A.

This unexpected behavior may be due in part to the fact that
the $T$=1, $T$=0 splitting of the $0^{-}$ states is sensitive
more to the long-range behavior of the tensor interaction
while the D-state admixture is a result of the tensor force also at
smaller ranges.
To this end, we note that the $\pi NN$ coupling constant
$g_{\pi}^{2}/(4\pi)$ is 14.9 for potential A, while it is 14.6 for OBEP C.
The larger D-state probability obtained for potential C is mainly due
to the larger cut-off mass (1.3 GeV as compared to 1.05 GeV for
potential A). The potential C, however, also shows a less repulsive
central isovector component. This is due to the larger coupling
constant for the $\delta$ meson and can also be deduced from a local
parameterization of these potentials \cite{piotr}.

Note that both the Bonn A and Bonn C results at the $1\hbar\omega$
$1p$-$1h$ level with $m^{*}$=939 MeV are larger than experiment.
The inclusion of the $3\hbar\omega$ $3p$-$3h$ configurations make
the splitting slightly {\it larger}. Only after the $3\hbar\omega$
$2p$-$2h$ configurations are allowed, do we get
a much smaller splitting, in agreement with experiment.
For Bonn A with $m^*$=$m$, the splitting
changes from 3.08 MeV for $1p$-$1h$ to 1.87 MeV for
$1p$-$1h$+$2p$-$2h$+$3p$-$3h$. The results for Bonn C with $m^*$=$m$
are similar ($2.75 \rightarrow 1.65$).

In previous studies we have seen that the relativistic features of the
DBHF approach, considering a density dependence of the Dirac spinors
for the nucleons, have a sizable effect on the NN interaction and
therefore modify the results of shell model calculations. These
effects may be characterized by an effective mass $m^{*}$ less
than the free nucleon mass $m$ \cite{walecka}.
We see from Table II that in the
$1p$-$1h$ calculation, the value of $\Delta E$ becomes smaller
as $(m^{*}/m)$ becomes smaller. In previous discussions, Zheng,
Zamick, and M\"uther \cite{zzm}
noted that the use of a Dirac effective mass $m^{*}$ less
than $m$ does not alter the tensor interaction much,
while it does enhance the spin-orbit interaction in
the nucleus by about a factor of $(m/m^*)$. From Table I, we see that at the
$1p$-$1h$ level, the full schematic interaction ($x$=$y$=1) gives a smaller
$\Delta E$ than the schematic central plus tensor interaction
($x$=0, $y$=1). Therefore the slight decrease observed in value of $\Delta E$,
if we consider $m^{*}$ less than $m$, is in line with the effect of the
spin-orbit interaction observed for the schematic potential. In
general, however, the effect of the density dependent Dirac spinors is
rather weak for the excitations, which we are investigating here.

We now return to the question of why the inclusion of $2p$-$2h$
causes $\Delta E$ to decrease.
A comparison of ``$1p$-$1h$+$3p$-$3h$'' and
``$1p$-$1h$+$2p$-$2h$+$3p$-$3h$'' results in Tables I and II shows
that the dominant configuration which leads to
the narrowing of the splitting is the $2p$-$2h$ state loosely described as
$0^{-}$($1p$-$1h$)$M^{+}$, where $M^{+}$ is a monopole
excitation. At the $2\hbar\omega$ $1p$-$1h$ level, the monopole state in
$^{16}\mbox{O}$ is an admixture of three configurations
$(1s_{1/2}\; 0s_{1/2}^{-1})$,
$(1p_{3/2}\; 0p_{3/2}^{-1})$, and
$(1p_{1/2}\; 0p_{1/2}^{-1})$.
Furthermore, when we repeat the $3\hbar\omega$ calculation in a smaller
space ($0s$, $0p$, $1s$-$0d$), so that only the configuration
$(1s_{1/2}\; 0s_{1/2}^{-1})$ enters, we obtain
almost the same small result for the splitting $\Delta E$.
This shows that it is mainly the $(1s_{1/2}\; 0s_{1/2}^{-1})$
component that is causing the decrease of $\Delta E$.

Some time ago, Zamick \cite{zamick} showed,
for {\it centroids} of particle-hole states, that
in perturbation theory,
it is the {\it isovector monopole} phonon-exchange between
two nucleons which causes the splitting between $T$=1 and $T$=0
states to decrease. The isoscalar phonon-exchange shifts the
$T$=1 and $T$=0 particle-hole centroids up by the same
amount thus giving no contribution to the splitting
$\Delta E$. On the other hand,
an isovector monopole exchange causes the centroid
of the $T$=0 particle-hole states
to get pushed up in energy while the centroid of the
$T$=1 states comes down. The net effect is to make the splitting
between the $T$=1 and $T$=0 states smaller. This effect has been
referred to as
the ``Zamick disease'' by Miller {\it et al.} \cite{mill}.

Obviously, our calculations in the large model-space reproduce this
effect, which, in terms of a renormalized interaction, would be described
as an exchange of an isovector monopole phonon. With inclusion of this
effect we get a rather good agreement with the experimental data for the energy
splitting between the $0^-$ excitations for $T$=1 and $T$=0.

Our predictions for the excitation energies  are all above the
experimental results. Even if we include the effects of $2p$-$2h$
configurations, our results are about 3 MeV or more above the experimental
data. As we have just discussed before, the centroid energy is
sensitive to the isoscalar component of the interaction; the same part
of the interaction is also responsible for the gross features of
the ground-state for the nucleus under consideration. It is known,
however, that the BHF approximation tends to predict too small radii
\cite{mmb}. Therefore we must also expect too large excitation energies
for the centroids. It is encouraging to observe that the DBHF effects,
characterized by an effective mass $m^*$ less than $m$, which improve
the results for the ground state calculations \cite{mmb} also reduce
the discrepancy between theory and experiments for the centroids of the
excitation energy. Here, we would like to recall that the results
presented here have been obtained from a realistic interaction without
any adjustable parameter or any single-particle energies taken from
other experimental data. In view of this fact we would call the agreement
between calculation and data satisfying even for the centroids of the
states.

Recently large basis shell model calculations for $psd$ shell nuclei
have been performed by Haxton and Johnson \cite{hax}
and by Warburton, Brown, and
Millener \cite{warb}. They note that several issues have to be resolved
to do these calculations correctly, i.e.,
one cannot just perform them blindly (see also \cite{millener}).
Part of the focus in these works is to handle large enough spaces,
so that the highly deformed states come down to respectably low energies.
The above groups, however, suppress the ``single-particle
excitations through two major shells'' --- this would exclude the
monopole excitations that we have emphasized above --- ``except
as they are needed to insure the removal of spuriousity''.
Thus on this point their approach is orthogonal to ours.

\section{Conclusions}

\noindent
We have demonstrated that the inclusion of $2 \hbar \omega$ monopole
excitations can be very important for the microscopic understanding of
nuclear excitations within the shell model. This can either be done by
renormalizing the effective interaction or, as has been done in the
present work, by enlarging the shell model space to include such
excitations explicitly. Studying the energy splitting between the
$J^{\pi}$=$0^-$
states with isospin $T$=0 and $T$=1, one finds that a reasonable
agreement can be obtained if such monopole excitations are taken into
account. Ignoring these configurations by performing $1p$-$1h$
calculations, one obtains an energy splitting which is too large by
about 50\% . This can be corrected by renormalizing the isovector part
of the NN interaction either empirically by reducing the tensor force
or by a perturbative calculation accounting for the exchange of
isovector monopole phonons \cite{zamick}. This also shows that an
effective interaction derived from nuclear structure calculations,
restricted to $1p$-$1h$ excitations would underestimate the tensor
force as compared to a realistic NN interaction.

In this work, we have presented a mechanism which causes the tensor
interaction in a nucleus to appear to be weaker than it actually is.
There are other formulations, in which the tensor interaction
really is weaker in the nuclear medium than in the free space. An example of
this is the universal scaling model of
Brown and Rho \cite{brown} and the application of this to finite nuclei by
Hosaka and Toki \cite{hsk}. In this model, not only the nucleon but also
the mesons, except for the pions, are less massive inside the nucleus.
In particular, since the $\rho$ meson is less massive,
the range of the nucleon-nucleon interaction due to $\rho$-exchange is
longer. Since the $\rho$-exchange gives a repulsive
contribution to the tensor interaction, making the range longer
will cause more cancellation with the attractive contributions due
to the $\pi$-exchange.

Clearly in our calculation, we get a weakening of the
tensor interaction from a much less exotic mechanism --- non
RPA core polarization --- which can equally well be described as configuration
mixing. Whether or not there is a phenomenological need for both
our mechanism and the universal scaling mechanism remains
to be seen. We have carefully studied one example: The
$J^{\pi}$=$0^-$ states in $^{16}\mbox{O}$. We will in the near
future have to look at many other cases with equal care.
In this regard, forthcoming work by P. Czerski {\it et al.} \cite{piotr}
should be of interest.
We also cite the experimental work of Hintz's group
\cite{hintz1,hintz2} on inelastic electron and proton scattering
to high spin ``stretch'' states in $^{208}\mbox{Pb}$ and also
work by E.J. Stepherson {\it et al.} on inelastic
proton scattering to $J^{\pi}$=$4^-$ stretch states in
$^{16}\mbox{O}$ \cite{stepherson}. Both groups conclude that
a considerable medium modification is needed to explain
the data.

\section*{Acknowledgment}
We thank B. R. Barrett, J. P. Vary and D. J. Millener
for useful discussions and for the critical reading of the
manuscript. We also thank R. Machleidt for
advice and encouragement. This work was partially
supported by the National Science Foundation, grant \#
PHY-9103011 and by the Department of Energy, Grant \# DE-FG05-86ER-40299.
One of us (D.C.Z) is grateful to the National Institute
for Nuclear Theory at the
University of Washington for its support during his visit.

\vspace{0.2in}

\begin{small}

\end{small}

\pagebreak

{\bf Table I}. The energies of
the $J^{\pi}$=$0^{-}$, $T$=0 and 1 states
in ${}^{16}\mbox{O}$ and their splitting.
Three kinds of configurations are used: (1) $1\hbar\omega$ $1p$--$1h$;
(2) $1\hbar\omega$ $1p$--$1h$ + $3\hbar\omega$ $3p$--$3h$
with the extra 2 holes in the $0p$ shell and
the extra 2 particles in the $sd$ shell;
(3) $1\hbar\omega$ $1p$--$1h$ + $3\hbar\omega$ $2p$--$2h$
+ $3\hbar\omega$ $3p$--$3h$.
Schematic interactions are used, as described in the text
[see Eq.(\ref{eq3})].

\begin{center}

\begin{small}
\begin{tabular}{l|ccc|ccc|ccc} \hline\hline
Interaction & \multicolumn{3}{c|}{$1p$--$1h$} &
\multicolumn{3}{c|}{$1p$--$1h$ + $3p$--$3h^{a)}$} &
\multicolumn{3}{c}{$1p$--$1h$ + $2p$--$2h$ + $3p$--$3h$} \\
 & $E_{0^{-},1}$ & $E_{0^{-},0}$ & $\Delta E$
 & $E_{0^{-},1}$ & $E_{0^{-},0}$ & $\Delta E$
 & $E_{0^{-},1}$ & $E_{0^{-},0}$ & $\Delta E$ \\ \hline
C      &19.953 &19.936 &0.017 &21.698 &21.733 &-0.035 &21.624 &21.515 &0.109 \\
C+T    &20.104 &16.960 &3.144 &22.561 &19.517 & 3.044 &21.428 &19.190 &2.238 \\
                                                        \hline
C+SO   &16.670 &16.624 &0.046 &18.674 &18.710 &-0.036 &17.607 &17.451 &0.156 \\
C+SO+T &16.792 &13.860 &2.932 &19.516 &16.616 & 2.900 &17.608 &15.606 &2.002\\
                                                 \hline\hline
\end{tabular}
$^{a)}$ When we add $3p$-$3h$ to the $J^{\pi}$=$0^-$ states, we simultaneously
add $2p$-$2h$ admixtures into the ground state.
\end{small}
\end{center}

\vspace{1.0in}

{\bf Table II}. Same as Table I but using the OBE potentials A and C
defined in \cite{mach} with an effective nucleon mass $m^*$.

\begin{center}

\begin{small}
\begin{tabular}{r|ccc|ccc|ccc} \hline\hline
Interaction & \multicolumn{3}{c|}{$1p$--$1h$} &
\multicolumn{3}{c|}{$1p$--$1h$ + $3p$--$3h$} &
\multicolumn{3}{c}{$1p$--$1h$ + $2p$--$2h$ + $3p$--$3h$} \\
Bonn & $E_{0^{-},1}$ & $E_{0^{-},0}$ & $\Delta E$
         & $E_{0^{-},1}$ & $E_{0^{-},0}$ & $\Delta E$
         & $E_{0^{-},1}$ & $E_{0^{-},0}$ & $\Delta E$ \\ \hline
A \hspace{0.05in} $m^{*}$=939
&17.187 &14.106 &3.08 &19.156 &15.952 &3.20 &17.036 &15.169 &1.87\\
         $m^{*}$=729
&17.037 &14.297 &2.74 &19.044 &16.201 &2.84 &16.121 &14.481 &1.64\\ \hline
C \hspace{0.05in} $m^{*}$=939
&16.940 &14.224 &2.72 &18.879 &16.014 &2.87 &16.233 &14.587 &1.65\\
         $m^{*}$=729
&16.715 &14.315 &2.40 &18.680 &16.150 &2.53 &15.277 &13.787 &1.49\\
                         \hline\hline
\end{tabular}
\end{small}
\end{center}

\end{document}